\documentclass[a4paper,11pt]{article}
\usepackage{jcappub} 
\usepackage{lineno}

\usepackage{graphicx,setspace}
\usepackage{dcolumn}
\usepackage{bm,url}

\usepackage{mathrsfs}
\usepackage{xcolor}
\usepackage{float}
\usepackage[normalem]{ulem}
\usepackage[utf8]{inputenc}
\usepackage{tensor}
\usepackage{physics}
\usepackage{blindtext}
\usepackage{slashed}
\usepackage{tikz}
\usepackage{amsthm,amsfonts,epsfig,natbib,amssymb}
\usepackage{tensor}
\usepackage{mathtools}
\usepackage[scaled]{beramono}
\usepackage[T1]{fontenc}
\usetikzlibrary{decorations.pathmorphing}

\title{\boldmath Notes on emergent conformal symmetry for black holes}

\author[a]{Ye-Sheng Xue}
\author[b]{Jie Jiang}
\author[a,c]{Ming Zhang}
\emailAdd{mingzhang@jxnu.edu.cn}
\affiliation[a]{Department of Physics, Jiangxi Normal University,\\ Nanchang 330022, China}
\affiliation[b]{Faculty of Arts and Sciences, Beijing Normal University,\\ Zhuhai 519087, China}
\affiliation[c]{Department of Physics and Astronomy, University of Waterloo,\\
Waterloo, Ontario N2L 3G1, Canada}

\abstract{We examine the motion of the massless scalar field and nearly bound null geodesics in the near-ring region of a black hole, which may possess either acceleration or a gravitomagnetic mass. Around such black holes, the photon ring deviates from the equatorial plane. In the large angular momentum limit, we demonstrate that the massless scalar field exhibits an emergent conformal symmetry in this near-ring region. Additionally, in the nearly bound limit, we observe the emergence of a conformal symmetry for the null geodesics that constitute the photon ring in the black hole image. These findings suggest that the hidden conformal symmetry, associated with the Lie algebra $\mathfrak{s l}(2, \mathbb{R})$,  persists even for black holes lacking north-south reflection symmetry, thereby broadening the foundation of photon ring holography. Finally, we show that the conformal symmetry also emerges for nearly bound timelike geodesics and scalar fields in proximity to the particle ring, and with specific mass around a Schwarzschild black hole.
}

\begin{document}
\maketitle
\flushbottom

\section{Introduction}

Photon rings associated with black holes serve important purposes from multiple perspectives. Firstly, they are closely linked to black hole images, as demonstrated by recent observations of the shadows of M87* \cite{EventHorizonTelescope:2019dse} and Sgr A* \cite{EventHorizonTelescope:2022wkp} by the Event Horizon Telescope. Secondly, photon rings are connected to gravitational waves as they control the photon ring quasinormal modes (QNMs) \cite{Cardoso:2017soq}. In a recent study \cite{Raffaelli:2021gzh}, it was discovered that an  emergent conformal symmetry associated with the $\mathfrak{s l}(2, \mathbb{R})$ Lie algebra (denoted as $\mathfrak{s l}(2, \mathbb{R})_{\mathrm{QNM}}$ symmetry) exists for the massless scalar field near the photon ring surrounding static and spherically symmetric black holes. This symmetry enables the recovery of weakly damped QNMs in the large angular momentum limit (eikonal limit) through the $\mathfrak{s l}(2, \mathbb{R})$ Lie algebra and its representations. The intricate self-similar structure of the photon region has also been found to be applicable to rotating Kerr black holes \cite{Hadar:2022xag}, suggesting that this emergent conformal symmetry allows for a holographic interpretation of microstates that are dual to Kerr black holes. Additionally, in \cite{Kapec:2022dvc}, the utilization of a self-dual warped three-dimensional Anti-de Sitter (AdS)  geometry illustrated a direct connection between the $\mathfrak{s l}(2, \mathbb{R})$ conformal symmetry of the photon ring (denoted as $\mathfrak{s l}(2, \mathbb{R})_{\mathrm{PR}}$ symmetry) around a near-extremal Kerr black hole and the $\mathfrak{s l}(2, \mathbb{R})$ isometries of the spacetime (denoted as $\mathfrak{s l}(2, \mathbb{R})_{\mathrm{ISO}}$). Along this line, \cite{Chen:2022fpl} made a novel discovery regarding the construction of the near-ring $\mathfrak{s l}(2, \mathbb{R})$ symmetry for warped AdS black holes. It was found that the highest-weight representations of this symmetry correspond to the eikonal QNM family. The $\mathfrak{s l}(2, \mathbb{R})_{\mathrm{QNM}}$ symmetry can be connected to the spacetime $\mathfrak{s l}(2, \mathbb{R})_{\mathrm{ISO}}$ in an another manner, as both emergent conformal symmetries are equivalent for the near-ring eikonal QNMs. The $\mathfrak{s l}(2, \mathbb{R})$ algebra was also investigated in \cite{Hadar:2022xag,Kapec:2022dvc,Chen:2022fpl} in relation to nearly bound null geodesics, studying the connection between the $\mathfrak{s l}(2, \mathbb{R})_{\mathrm{PR}}$ symmetry for photons and the $\mathfrak{s l}(2, \mathbb{R})_{\mathrm{QNM}}$ symmetry for QNMs. Recently, by utilizing the $\mathfrak{s l}(2, \mathbb{R})_{\mathrm{ISO}}$, the QNMs for scalar, vector, and spinor fields were obtained for the self-dual warped three-dimensional AdS black hole \cite{Chen:2023zvd}. In line with the aforementioned studies, \cite{Hashimoto:2023buz,Riojas:2023pew} discovered a peculiar form of the scalar QNM spectra near the photon sphere of a Schwarzschild black hole in an asymptotically AdS spacetime. The imaginary part of the QNMs was found to be determined by the photon sphere, implying the existence of a photon-sphere subsector in the dual holographic thermal conformal field theory (CFT).

Previous studies have primarily focused on exploring the emergence of $\mathfrak{s l}(2, \mathbb{R})$ conformal symmetry in the dynamics of the massless scalar field and the null geodesics around black holes that possess $\mathbb{Z}_2$ symmetry (i.e., the north-south reflection symmetry). In this setup, the closed photon rings are located on the equatorial plane, which is unstable in the radial direction but stable in the latitudinal direction. In the eikonal limit, the radial location at which the maximum of the effective potential for photons occurs is almost the same as that for the massless scalar field. 

Symmetry is a fundamental concept in physics, manifesting in various forms in different contexts. A well-known example is CPT (which stand for charge conjugation, parity, and time reversal, respectively) invariance in the Standard Model \cite{Luders:1957bpq}. However, symmetries can also be broken, as evidenced by the violation of parity \cite{Lee:1956qn,Wu:1957my}. When we state that a black hole solution exhibits $\mathbb{Z}_2$ symmetry, we mean that the metric remains invariant under the transformation $(\theta, \phi) \rightarrow(\pi-\theta, \phi)$. Consequently, the horizon of the corresponding black hole is isometric across its two hemispheres, connected by the equatorial plane, the fixed point of the $\mathbb{Z}_2$ symmetry \cite{Cunha:2018uzc}. Interestingly, the $\mathbb{Z}_2$ asymmetry (i.e., reflection asymmetry on the equatorial plane) of black hole spacetime, including its horizon and photon rings, can arise if Chern-Simons parity-violating terms are incorporated into the low-energy effective field theory \cite{Cano:2019ore,Tahara:2023pyg}. The breaking of $\mathbb{Z}_2$ symmetry in slowly-spinning black holes was also discovered in the effective field theory \cite{Cardoso:2018ptl}. This led to the detection of non-zero Tidal Love Numbers (TLNs), which qualitatively describe the tidal deformation of an extended object in a gravitational field. This finding is noteworthy because in classical general relativity, the TLNs for black holes are zero \cite{Binnington:2009bb,Cardoso:2017cfl}. In a separate study \cite{Cunha:2018uzc}, a black hole without $\mathbb{Z}_2$ symmetry was identified in a gravitational theory non-minimally coupled with scalar fields. In this case, the horizon of the black hole is $\mathbb{Z}_2$ odd.

The absence of $\mathbb{Z}_2$ symmetry can have significant phenomenological impacts, influencing light lensing as well as light absorption and scattering. The violation of $\mathbb{Z}_2$ symmetry in a black hole could potentially be probed in astrophysical scenarios through gravitational waves emanating from compact binary coalescences \cite{Seto:2007tn,Yagi:2017zhb,Zhao:2019xmm,Jenks:2023pmk}. However, despite these potential effects, the shadow of a black hole has been shown to retain $\mathbb{Z}_2$ symmetry \cite{Cunha:2018uzc}. As a result, the exact implications of the absence of $\mathbb{Z}_2$ symmetry remain uncertain. Given these observations, it's crucial to investigate whether the $\mathfrak{s l}(2, \mathbb{R})$ emergent conformal symmetry persists in black hole spacetime in the absence of $\mathbb{Z}_2$ symmetry.

There are two well-known types of black holes that lack $\mathbb{Z}_2$ symmetry in classical general relativity: the accelerating black hole and the Taub-NUT (Newman-Unti-Tamburino) black hole. Both belong to the Plebanski-Demianski family of solutions in Einstein gravity \cite{Plebanski:1976gy,Podolsky:2021zwr,Podolsky:2022xxd,Chng:2006gh,Barrientos:2023tqb}. The accelerating black hole is characterized by a cosmic string that connects the poles with the spacetime boundary, resulting from the presence of an acceleration parameter. The Taub-NUT black hole possesses gravitomagnetic mass (or NUT charge, Misner charge) and exhibits string singularities on the poles that extend to spatial infinity. In our previous related investigations \cite{Zhang:2020xub,Zhang:2021pvx} (see also \cite{Gibbons:2016isj,Lim:2020bdj,Frost:2020zcy} for the accelerating black hole), it was found that the acceleration and the gravitomagnetic mass cause the photon ring of the black hole to deviate from the equatorial plane. These off-equatorial photon rings have a constant radius and latitude. Remarkably, the contour of the shadow still retains the $\mathbb{Z}_2$ symmetry.  On the other hand, the behavior of the scalar field in the vicinity of these black holes has been studied, notably in \cite{Fontana:2022whx} for the accelerating AdS black hole and in \cite{Kalamakis:2020aaj} for the Taub-NUT-AdS black hole. However, the dynamics of the scalar field in the near-photon-ring region of a $\mathbb{Z}_2$-asymmetric spacetime are not yet fully understood. It is unclear whether the emergent $\mathfrak{s l}(2, \mathbb{R})$ conformal symmetry, which has been observed in other contexts, persists in these cases. This question arises from the observation that the effective potentials for photons in these $\mathbb{Z}_2$-asymmetric spacetimes also lack $\mathbb{Z}_2$ symmetry, as the photons do not remain confined to the equatorial planes. Simultaneously, the motion of massless scalar fields in the radial direction is coupled with their motion in the latitudinal direction in these spacetimes. This complex situation is distinct from previous cases involving black holes with $\mathbb{Z}_2$ symmetry. In those instances, both the effective potentials of the photons and the scalar field are independent of $\theta$, the polar angle. This stark contrast underscores the need for further investigation to fully comprehend the implications of $\mathbb{Z}_2$ asymmetry in black hole spacetimes.

Another problem we aim to investigate is the existence of conformal symmetry near the particle ring. This involves examining the characteristics of the equations of motion for both the massive particle and the massive scalar field in the vicinity of the black hole. Considering that there are both stable and unstable circular orbits for massive particles around black holes, it is intriguing to explore the behavior of the massive scalar field around these two types of orbits. Studying the emergent conformal symmetry for black holes without $\mathbb{Z}_2$ symmetry will provide useful insights for our investigation of emergent conformal symmetry around the particle ring.

In this paper, we present our findings regarding the existence and preservation of the emergent $\mathfrak{s l}(2, \mathbb{R})_{\mathrm{PR}}$ and $\mathfrak{s l}(2, \mathbb{R})_{\mathrm{QNM}}$ conformal symmetry in black hole spacetimes that lack $\mathbb{Z}_2$ symmetry, as well as around the particle rings of a black hole. Specifically, we focus on the accelerating Schwarzschild black hole in an AdS background in section \ref{hasbh} and the Taub-NUT black hole in an AdS background in section \ref{hctnbh}. Our investigations and results also extend to the case of a flat background. For each scenario, we start by identifying the location of the photon ring, followed by an examination of the dynamics of the massless scalar field in the region near the photon ring. We also explore the phase space of the nearly bound null geodesic close to the photon ring. In section \ref{kjhj4857}, we study the emergent conformal symmetry of the massive particle and massive scalar field near the particle ring around the Schwarzschild black hole. Finally, in section \ref{seccr}, we conclude that the emergent $\mathfrak{s l}(2, \mathbb{R})_{\mathrm{PR}}$ and $\mathfrak{s l}(2, \mathbb{R})_{\mathrm{QNM}}$ conformal symmetry remains intact in black hole spacetimes without $\mathbb{Z}_2$ symmetry. We also conclude that the $\mathfrak{s l}(2, \mathbb{R})$ algebra applies to the massive particle and the  massive scalar field near the particle ring of the Schwarzschild black hole. It's important to note that in sections \ref{hasbh}, \ref{hctnbh}, and \ref{kjhj4857}, the symbols we use to represent most of the physical quantities, such as the effective potentials of particles and scalar fields, are specific to and only applicable within their respective sections.

%%%%%%%%%%%%%%%%%%%%%%%%%%%%%%%%%%%%%%%%%%%%%%%%%%%%
%%%%%%%%%%%%%%%%%%%%%%%%%%%%%%%%%%%%%%%%%%%%%%%%%%%%
%%%%%%%%%%%%%%%%%%%%%%%%%%%%%%%%%%%%%%%%%%%%%%%%%%%%
%%%%%%%%%%%%%%%%%%%%%%%%%%%%%%%%%%%%%%%%%%%%%%%%%%%%
%%%%%%%%%%%%%%%%%%%%%%%%%%%%%%%%%%%%%%%%%%%%%%%%%%%%

\section{Notes on emergent conformal symmetry near off-equatorial photon ring}

\subsection{Emergent  conformal symmetry for accelerating black hole}\label{hasbh}

We will investigate the emergent conformal symmetry of the massless scalar field and the behavior of the nearly bound null geodesic around an accelerating black hole. Specifically, we will focus on the accelerating Schwarzschild AdS black hole. However, it is important to note that the findings can be extrapolated to the accelerating black hole without a cosmological constant. A critical observation is that the acceleration of the black hole causes a loss of the $\mathbb{Z}_2$ symmetry typically exhibited by most black holes. In the case of the accelerating Schwarzschild black hole, the symmetry is no longer spherically symmetric, resulting in the deviation of the photon ring from the equatorial plane, as we will show in the following analysis.

\subsubsection{Off-equatorial photon ring}

The metric of the accelerating AdS black hole can be expressed as
\begin{equation}
\label{met}ds^2=\frac{1}{\Omega^2}\left[-\frac{f(r)}{\alpha^2} \mathrm{d}t^2 +\frac{\mathrm{d}r^2}{f(r)} +r^2\left(\frac{\mathrm{d}\theta^2}{P(\theta)}+P(\theta) \sin^2 \theta\frac{\mathrm{d}\phi^2}{K^2}\right)\right],
\end{equation}
where
\begin{align}
\Omega&=1-Ar\cos\theta,\\P(\theta)&=1-2mA\cos\theta,\label{macie38}\\f(r)&=\left(1-A^2 r^2\right)\left(1-\frac{2m}{r}\right)+\frac{r^2}{l^2}.
\end{align}
When the acceleration parameter $A$ is set to zero, the metric reduces to the Schwarzschild AdS form. The parameters $l$ and $m$ correspond to the AdS radius and mass parameter respectively. $l$ is related with the negative cosmological constant $\Lambda$ by $\Lambda=-3/l^2$. The conformal boundary of the spacetime, determined by $\Omega=0$, is given by $r_b=1/A\cos\theta$. The factor $\alpha=\sqrt{1-A^2 l^2}$ serves to adjust the time coordinate for an asymptotic observer \cite{Anabalon:2018ydc,Anabalon:2018qfv}. Additionally, the parameter $K\neq 1$ is introduced to account for conical deficits at the poles of the black hole. This restricts $\phi$ to the interval $[0, 2\pi K)$. Selecting $K=K_+\equiv P(\theta=0)=1-2m A$ eliminates the conical deficit at the north pole. Consequently, the conical deficit at the south pole can be expressed as
\begin{equation}
\delta =2\pi\left(1-\lim_{\theta\rightarrow\pi}\frac{1}{\theta}\sqrt{\frac{g_{\phi\phi}}{g_{\theta\theta}}}\right)=2\pi\left(1-\frac{P_{-}}{K_{+}}\right)=-\frac{8\pi m A}{1-2m A},
\end{equation}
where $P_-=P(\theta=\pi)$. The cosmic string tension $\mu$ is related to the conical deficit by $\mu=\delta/8\pi$ \cite{Appels:2017xoe}.

The null geodesic equations for the photon in the accelerating Schwarzschild AdS black hole spacetime have been previously investigated in \cite{Grenzebach:2015oea,Zhang:2020xub,Lim:2020bdj}. However, these studies did not include the parameters $\alpha$ and $K$ in the equations of motion for the particles. \footnote{While this may not be particularly significant in the kinematic and dynamic analyses, incorporating these parameters becomes important for topics related to thermodynamics and field theory.} We will adapt their findings by including the parameters $\alpha$ and $K$.

The Lagrangian describing the photon in the spacetime \eqref{met} is given by
\begin{equation}\label{lagc}
\mathcal{L}=-\frac{1}{2} g_{\mu \nu} \dot{x}^{\mu} \dot{x}^{\nu}=0,
\end{equation}
where the dot represents the derivative with respect to the affine parameter $\tau_0$. Since $\xi_t$ and $\xi_\phi$ are Killing vectors of the spacetime, we have two conserved quantities, namely energy $E$ and orbital angular momentum $S$, which are conjugate momenta to $\dot{t}$ and $\dot{\phi}$ as
\begin{equation}\label{ptpphi}
P_t =\frac{\partial \mathcal{L}}{\partial \dot{t}}=E, \quad  P_\phi =-\frac{\partial \mathcal{L}}{\partial \dot{\phi}}=S.
\end{equation}
Combining equations \eqref{lagc} and \eqref{ptpphi}, we obtain
\begin{equation}\label{wthoas}
-\frac{\alpha ^2 E^2}{f(r)}+\frac{\dot{r}^2}{\Omega ^4 f(r)}+\frac{K^2 S^2 \csc ^2\theta }{r^2 P(\theta )}+\frac{\dot{\theta }^2 r^2}{\Omega ^4 P(\theta )}=0.
\end{equation}
By introducing a separation constant $Z$, \eqref{wthoas} can be split into the radial and angular parts as
\begin{equation}\label{rdthdz}
\frac{\dot{r}^2}{\Omega ^4 f(r)}-\frac{\alpha ^2 E^2}{f(r)}=-\frac{K^2 S^2 \csc ^2\theta }{r^2 P(\theta )}-\frac{\dot{\theta }^2 r^2}{\Omega ^4 P(\theta )}=Z.
\end{equation}
Combining \eqref{ptpphi} and \eqref{rdthdz}, we derive the equations of motion for the null geodesic of the photon in the accelerating AdS black hole as 
\begin{align}
\frac{r^2}{\Omega^2} \frac{\mathrm{d}t}{\mathrm{d} \tau}&=\frac{ r^2 E}{f},\\\frac{r^2}{\Omega^2} \frac{\mathrm{d} \phi}{\mathrm{d} \tau}&=\frac{K^2 S}{P \sin ^2 \theta},\label{dphidtau}\\\left(\frac{r^2}{\Omega^2}\right)^2\left(\frac{\mathrm{d} \theta}{\mathrm{d} \tau}\right)^2&=-\frac{K^2 S^2}{\sin^2 \theta}-Z P r^2 \equiv \Theta(\theta),\label{dthetadtau}\\\left(\frac{r^2}{\Omega^2}\right)^2\left(\frac{\mathrm{d} r}{\mathrm{d} \tau}\right)^2&=\alpha^2 r^4 E^2+r^4 f Z \equiv R(r),\label{radrr}\end{align}
where we have defined the functions $\Theta(\theta)$ and $R(r)$ relating with the latitudinal and radial components of the four-velocity, respectively. It is worth noting that when the equations are simplified to the case without a cosmological constant, the third and fourth equations may differ superficially from those in \cite{Frost:2020zcy} due to the different means of variable separation in \eqref{wthoas}. However, the description of the motion for photons remains unchanged.

The radius and latitude of the photon ring are denoted as $r_p$ and $\theta_p$ respectively. For the circular orbit of the photon around the accelerating AdS black hole in the domain of outer communication, where $f(r)>0$ and $g(\theta)>0$, the following conditions must be satisfied:
\begin{equation}\label{poc2}
\Theta\left(\theta_p\right)=0, \quad \dot{\Theta}\left(\theta_p\right)=0, \quad \ddot{\Theta}\left(\theta_p\right)<0;
\end{equation}
\begin{equation}\label{poc1}
R\left(r_p\right)=0, \quad R^{\prime}\left(r_p\right)=0, \quad R^{\prime \prime}\left(r_p\right)>0.
\end{equation}
The dot and prime notations indicate derivatives with respect to $\theta$ and $r$ respectively. The condition $\ddot{\Theta}\left(\theta_p\right)<0$ indicates latitudinal stability, while $R^{\prime \prime}\left(r_p\right)>0$ indicates radial instability for the photon on the light rings. After some algebraic calculations, we obtain the radius $r_p$, latitudinal angle $\theta_p$, separation constant $Z_p$, and the impact parameter $\lambda_p$ defined as the ratio of angular momentum $S_p$ to energy $E_p$ for the photon ring as
\begin{align}
r_p&=\frac{\sqrt{12 m^2 A^2+1}-1}{2 m A^2},\\
\theta_p&=\arccos\left(\frac{1-\sqrt{12 m^2 A^2+1}}{6 m A}\right),\label{csothetap}\\
Z_p&=\frac{54 A^2 K^2 m^2 S^2 \left(1-\sqrt{12 A^2 m^2+1}\right)}{r_p^2 \left(12 A^2 m^2+\sqrt{12 A^2 m^2+1}-1\right)^2},\\
\lambda_p &=\frac{S_p}{E_p}=\frac{\alpha  E l r_p \sin \theta _p \sqrt{P(\theta_p) r_p}}{K \sqrt{l^2 \left(A^2 r_p^2-1\right) \left(2 m-r_p\right)+r_p^3}}.\label{lwwhwit}
\end{align}
When $A\to 0$, the Schwarzschild case is recovered, with $r_p\to 3 m$, $\theta_p\to \pi/2$, $Z_p\to -S^2/9m^2$, and $\lambda_p=3\sqrt{3}m$. It is evident that the photon ring in the accelerating AdS black hole deviates from the equatorial plane.

The radial function $R(r)$ alone cannot determine the location of the photon ring, as indicated by \eqref{poc2} and \eqref{poc1}. To find an effective potential that directly yields the radius of the photon, we can express the radial component of the four-velocity, $\mathrm{d}r/\mathrm{d}\tau$, in a form independent of the separation constant $Z$,
\begin{equation}
\frac{\mathrm{d}r}{\mathrm{d}\tau}=\Omega^2 \sqrt{\mathcal{V}(r)},
\end{equation}
where the effective potential $\mathcal{V}(r)$ of the photon  is defined as
\begin{equation}\label{saadseff}
\mathcal{V}(r)=\alpha^2 E^2-\frac{K^2 S_p^2}{P(\theta_p) \sin^2 \theta_p} \frac{f(r)}{r^2}.
\end{equation}
Therefore, the energy $E_p$ and the radius $r_p$ of the photon ring can be determined by the condition
\begin{equation}
\mathcal{V} (r_p)=\mathcal{V}^\prime (r_p)=0.
\end{equation}
It is worth noting that in the vanishing acceleration limit, the effective potential reduces to the Schwarzschild case $h_s(r)=E^2-f(r)/r^2$. Another important thing to note is that the deformed effective potential $\mathcal{V}(r)$ is constructed after we solve the latitudinal angle $\theta_p$ of the photon ring using \eqref{poc2} and \eqref{poc1}.

%%%%%%%%%%%%%%%%%%%%%%%%%%%%%%%%%%%%%%%%%%%%%%%%%%%%
%%%%%%%%%%%%%%%%%%%%%%%%%%%%%%%%%%%%%%%%%%%%%%%%%%%%
%%%%%%%%%%%%%%%%%%%%%%%%%%%%%%%%%%%%%%%%%%%%%%%%%%%%
%%%%%%%%%%%%%%%%%%%%%%%%%%%%%%%%%%%%%%%%%%%%%%%%%%%%
%%%%%%%%%%%%%%%%%%%%%%%%%%%%%%%%%%%%%%%%%%%%%%%%%%%%

\subsubsection{\texorpdfstring{$\mathfrak{sl}(2,\mathbb{R})_{\text{QNM}}$}{lg2}  of massless scalar field}

We will derive the wave equation for the massless scalar field $\Phi$ non-minimally coupled to the Ricci scalar $\mathcal{R}$ in the accelerating AdS spacetime background \eqref{met}. The action for the scalar field is given by
\begin{equation}
\mathcal{S}_m=\int_{\mathcal{M}} d^4 x \sqrt{-g}\left(\partial_\mu \Phi \partial^\mu \Phi-\eta \mathcal{R} \Phi^2\right).
\end{equation}
To ensure the separation of the angular and radial equations, we choose the coupling constant $\eta=1/6$ so that the scalar field is conformally coupled to the spacetime curvature. Here, $\mathcal{R}$ represents the Ricci scalar of the spacetime \eqref{met}.

The equation of motion for the massless scalar field is given by
\begin{equation}\label{kgeqeom}
\nabla^\alpha \nabla_\alpha \Phi-\eta \mathcal{R} \Phi=0.
\end{equation}
To expand the first term on the left-hand side of \eqref{kgeqeom}, we directly use the metric \eqref{met}. For calculating the second term, we introduce a metric $\tilde{g}_{\alpha \beta}$ that is conformal to $g_{\alpha \beta}$ in \eqref{met}, defined as
\begin{equation}\label{eioroj8}
\tilde{g}_{ab}=\Omega^2 g_{ab}.
\end{equation}
Denoting the Ricci scalar of the conformal metric $\tilde{g}_{\alpha \beta}$ as $\mathcal{R}_{\tilde{g}}$, we have  \cite{wald2010general}
\begin{equation}
\mathcal{R}_{\tilde{g}}=  \Omega^{-2}\left[\mathcal{R}-6 g^{a c} \nabla_a \nabla_c \ln \Omega -6 g^{a c}\left(\nabla_a \ln \Omega\right) \nabla_c \ln \Omega\right].
\end{equation}
By substituting $\mathcal{R}$ with $\mathcal{R}_{\tilde{g}}$, we obtain
\begin{equation}
\begin{aligned}&\left(2 P \Omega_\theta^2-\frac{\partial_\theta(P \sin \theta) \Omega_\theta}{\sin \theta}-\frac{P \partial_{\theta \theta} \Omega}{\Omega}-\frac{R_{\tilde{g}} r^2}{6}-\partial_r\left(r^2 f\right) \Omega_r+2 f r^2 \Omega_r^2-\frac{f r^2 \partial_{r r} \Omega}{\Omega}\right) \Phi\\&+\partial_r\left(r^2 f\right) \partial_r \Phi -\frac{\alpha^2 r^2}{f} \partial_{t t} \Phi+r^2 f \partial_{r r} \Phi+\frac{K^2}{P \sin ^2 \theta} \partial_{\phi \phi} \Phi+\frac{\partial_\theta(P \sin \theta)}{\sin \theta} \partial_\theta \Phi \\&+P \partial_{\theta \theta} \Phi-2 r^2 f \Omega_r \partial_r \Phi-2 P \Omega_\theta \partial_\theta \Phi=0,
\end{aligned}
\end{equation}
where $\Omega_r\equiv\partial_r \Omega/\Omega$ and $\Omega_\theta\equiv\partial_\theta \Omega/\Omega$. The equation is almost identical to  (4.4) in \cite{Fontana:2022whx}, with the exception of the second and the fourth terms in the second line.

After multiplying the above equation by $\Omega^{-1}$, we decompose $\Phi$ as
\begin{equation}
\Phi=\frac{\Omega e^{-i \omega t}\psi(r) \zeta(\theta) e^{i m_0 \phi}}{r}.
\end{equation}
Here, $\omega=\omega_R-i \omega_I$ represents the complex frequency of the QNMs, and $m_0$ is a dimensionless number. We introduce tortoise-like coordinates
\begin{equation}
\mathrm{d} r_*=\frac{1}{f} \mathrm{d} r, \quad \mathrm{d} \Theta=\frac{1}{P \sin \theta} \mathrm{d} \theta.
\end{equation}
Using these coordinates, the radial equation and the angular equation can be written separately as
\begin{equation}\label{radiaeq}
\left(\frac{\partial^2}{\partial r_*^2}+\alpha ^2 \omega ^2-V(r)\right) \psi=0,\end{equation}\begin{equation}\label{angueq}\left(\frac{\partial^2}{\partial \Theta^2}-K^2 m_0^2+\vartheta(\theta)\right) \zeta=0,
\end{equation}
where the effective potential functions are defined as
\begin{equation}\label{radeff}
V(r)=f\left(\frac{\lambda}{r^2}-\frac{f}{3 r^2}+\frac{\partial_r f}{3 r}-\frac{\partial_{r r} f}{6}\right),
\end{equation}
\begin{equation}\label{thpiei49}
\vartheta(\theta)=P \sin ^2 \theta\left(\lambda-\frac{P}{3}+\frac{\cot \theta \partial_\theta P}{2}+\frac{\partial_{\theta \theta} P}{6}\right).
\end{equation}
Here, $\lambda$ is a separation constant, which is approximated as $\ell(\ell+1)+1/3$, where $\ell$ is the spherical harmonic index. We have $\phi\in [0,2 \pi K)$ at the poles, which implies $m_0=\bar{m} / K$, with $\bar{m} \in \mathbb{Z}$ representing the angular quantum number. Please note that if we do not utilize the conformal gauge given by \eqref{eioroj8}, and instead directly calculate $\mathcal{R}$ from \eqref{met}, we will not be able to separate the radial and angular parts of \eqref{kgeqeom}; readers can refer to \cite{Hawking:1997ia}.

Appropriate boundary conditions for the radial function $\psi$ are given by
\begin{equation}\label{boud1c}
\psi \sim \begin{cases}e^{-i \omega r_*}, & r_* \rightarrow-\infty\left(r \rightarrow r_{+}\right), \\ e^{i \omega r_*}, & r_* \rightarrow+\infty\left(r \rightarrow r_b\right),\end{cases}
\end{equation}
which ensure that there are no outgoing/ingoing waves at the event horizon $r_+$/acceleration horizons $r_b$ of the black hole. To ensure that the scalar field is finite at each pole, we impose the following boundary conditions in the angular directions:\begin{equation}\label{boud12c}
\zeta (\theta) \sim \begin{cases}e^{+|m_0| \Theta}, & \Theta \rightarrow-\infty(\theta \rightarrow 0), \\ e^{-|m_0| \Theta}, & \Theta \rightarrow+\infty(\theta \rightarrow \pi).\end{cases}
\end{equation}The QNM frequencies, which are complex values of $\omega$ satisfying \eqref{radiaeq}, \eqref{angueq} along with the boundary conditions \eqref{boud1c}, \eqref{boud12c}, can be solved numerically (see, for example, \cite{Destounis:2020pjk,Zhang:2023yco} for the case without a cosmological constant case; see also \cite{Lei:2023mqx} of solving the QNMs by using the B-cycle quantization condition in four-dimensional $\mathcal{N}=2$ supersymmetric gauge theories).  However, for the purpose of this discussion, we do not need to perform such calculations.

%%%%%%%%%%%%%%%%%%%%%%%%%%%%%%%%%%%%%%%%%%%%%%%%%%%%
%%%%%%%%%%%%%%%%%%%%%%%%%%%%%%%%%%%%%%%%%%%%%%%%%%%%
%%%%%%%%%%%%%%%%%%%%%%%%%%%%%%%%%%%%%%%%%%%%%%%%%%%%
%%%%%%%%%%%%%%%%%%%%%%%%%%%%%%%%%%%%%%%%%%%%%%%%%%%%
%%%%%%%%%%%%%%%%%%%%%%%%%%%%%%%%%%%%%%%%%%%%%%%%%%%%

In the eikonal limit $\lambda\gg 1$, the radial effective potential \eqref{radeff} simplifies to
\begin{equation}\label{effeik}
\lim_{\lambda\gg 1}V(r)\to V_\lambda (r)=\lambda\frac{f(r)}{r^2}.
\end{equation}
In this regime, we define
\begin{equation}
\mathcal{Q}_{\lambda \omega}\left(r_*(r)\right)=\alpha ^2 \omega ^2-V_\lambda (r).
\end{equation}
As a result, the radial equation \eqref{radiaeq} can be written as a Schrödinger-like equation
\begin{equation}\label{frscheq}
\frac{\mathrm{d}^2 \psi_{\lambda \omega}}{\mathrm{d} r_*^2}+\mathcal{Q}_{\lambda \omega}\left(r_*\right) \psi_{\lambda \omega}=0.
\end{equation}

Now we can analyze the behavior of the scalar field in the eikonal limit using \eqref{frscheq}, specifically around the photon ring determined by the effective potential \eqref{saadseff}. Remarkably, we find that
\begin{equation}
V_\lambda (r) \sim \mathcal{V}(r),
\end{equation}
which means that the location $r_0$ of the local maximum of $V_\lambda (r)$ coincides with the location $r_p$ of the photon ring. With this observation, we can expand $\mathcal{Q}_{\lambda \omega}\left(r_*(r)\right)$ around the photon ring $r_*(r_p)$ (or alternatively, the locally extremal point $r_0$ of $V_\lambda$) as
\begin{equation}\label{expqdklj898}
\mathcal{Q}_{\lambda \omega}\left(r_*\right) \approx \mathcal{Q}_0(\lambda, \omega, r_*(r_0))+\frac{1}{2} \mathcal{Q}_0^{(2)}(\lambda,r_*(r_0))\left[r_*(r)-r_*(r_0)\right]^2,
\end{equation}
Here, we have neglected  the higher-order terms and  we have also defined
\begin{equation}
\mathcal{Q}_0^{(n)}\equiv\left.\frac{\mathrm{d}^n \mathcal{Q}_{\lambda \omega}\left(r_*\right)}{\mathrm{d} r_*^n}\right|_{r_*(r)=r_*(r_0)}.
\end{equation}
Let us note here that  it would be impossible to judge whether  $ \mathcal{Q}_{\lambda \omega}\left(r_*\right)$ can be expanded as \eqref{expqdklj898} around the photon ring  if the effective potential of the photon is not written in form of \eqref{saadseff}, where the latitudinal angle $\theta_p$ is implemented.

Then we can introduce  definitions
\begin{align}\label{ghkd758}
x & =\left[2 \mathcal{Q}_0^{(2)}(\lambda)\right]^{1 / 4}\left[r_*-r_*(r_0)\right], \\h(\lambda, \omega) & =\frac{\mathcal{Q}_0(\lambda, \omega)}{\sqrt{2 \mathcal{Q}_0^{(2)}(\lambda)}}.
\end{align}
With them,  \eqref{radiaeq} can be rewritten as
\begin{equation}H \psi_{\lambda \omega}\left(r_*(x)\right)=h(\lambda, \omega) \psi_{\lambda \omega}\left(r_*(x)\right),
\end{equation}
where
\begin{equation}
H=-\frac{\mathrm{d}^2}{\mathrm{d} x^2}-\frac{1}{4} x^2
\end{equation}
can be interpreted as the Hamiltonian governing the dynamics of the massless scalar field.

Following the approach in \cite{Raffaelli:2021gzh}, we introduce the operators
\begin{align}
J_1 & =-\frac{i}{2}\left(x \frac{\mathrm{d}}{\mathrm{d} x}+\frac{1}{2}\right),\label{jop1} \\
J_2 & =\frac{i}{2}\left(\frac{\mathrm{d}^2}{\mathrm{d} x^2}-\frac{1}{4} x^2\right), \label{jop2}\\
J_3 & =-\frac{i}{2}\left(\frac{\mathrm{d}^2}{\mathrm{d} x^2}+\frac{1}{4} x^2\right)=\frac{i}{2} H, \label{jop3}\\
J_{ \pm}&= \pm i J_1-J_2.\label{jop4}
\end{align}
These operators satisfy the $\mathfrak{s l}(2, \mathbb{R})$ algebra
\begin{equation}\label{sl2r}
\left[J_{\pm}, J_{\mp}\right]=\mp 2 J_3, \quad\left[J_3, J_{ \pm}\right]= \pm J_{ \pm},
\end{equation}
to which we can refer as the $\mathfrak{s l}(2, \mathbb{R})_{\mathrm{QNM}}$ algebra, as done in \cite{Hadar:2022xag}. The eigenstates of $J_3$ are given by
\begin{equation}
J_3 \psi_h=h \psi_h,
\end{equation}
with $h=1/4, \, 3/4$ being the corresponding eigenvalues. This implies that the QNM frequencies can be described by two infinite representations of $\mathrm{SL}(2, \mathbb{R})$. The condition $J_+ \psi_{\lambda\omega}=0$, known as the highest-weight condition, satisfies the boundary condition \eqref{boud1c}, and the higher overtones can be seen as descendants of the $\mathfrak{s l}(2, \mathbb{R})$ algebra starting from the primary states with $h=1/4, 3/4$.

The emergence of the $\mathfrak{s l}(2, \mathbb{R})_{\mathrm{QNM}}$ conformal algebra for the scalar field, as identified in the context of the accelerating black hole background, hinges on two conditions. First, the scalar field must be conformally coupled to the spacetime curvature. This allows the radial and angular equations of motion for the scalar field to be separable. Second, the effective potential of the photon is expressed in the form of  \eqref{saadseff} as opposed to \eqref{radrr}. This form makes it apparent that the radial effective potential of the photon reaches an extremum at the radial location of the photon ring. However, it's important to underscore a unique aspect of this emerging $\mathfrak{s l}(2, \mathbb{R})_{\mathrm{QNM}}$ algebra for the scalar field, especially when compared to those found in \cite{Raffaelli:2021gzh,Hadar:2022xag,Kapec:2022dvc,Chen:2022fpl} for black holes with $\mathbb{Z}_2$ symmetry. In the $\mathbb{Z}_2$ symmetric cases, the scalar field is minimally coupled to the background geometry. In contrast, for the accelerating black hole scenario, the scalar field must be conformally coupled to the Ricci scalar.  Furthermore, in the $\mathbb{Z}_2$ symmetric cases, the latitudinal effective potential of the scalar field also exhibits $\mathbb{Z}_2$ symmetry (for instance, the latitudinal effective potential for the Schwarzschild black hole can be obtained by setting $P=1$ in \eqref{thpiei49}). However, in the absence of $\mathbb{Z}_2$ symmetry, the latitudinal effective potential loses this symmetry. Therefore, in a black hole spacetime without $\mathbb{Z}_2$ symmetry, the extrema of the radial effective potentials of the photon and the conformally coupled scalar field coincide, but the extrema of the latitudinal effective potentials of the photon and the scalar field do not. They only coincide in the limiting case of $A\to 0$, which corresponds to the Schwarzschild case.

In the limit of large spherical harmonic index $\ell$, corresponding to the eikonal limit $\lambda \gg 1$, \eqref{frscheq} can be solved analytically using the WKB (Wentzel–Kramers–Brillouin) method \cite{Iyer:1986np}. This gives the expression
\begin{equation}\label{wkb}
h(\lambda, \omega)=-i\left(n+\frac{1}{2}\right)+\mathcal{O}\left(\frac{1}{\lambda}\right),
\end{equation}
where the dimensionless nonnegative integer $n$ represents the overtone number. In this regime, as mentioned earlier, the radial effective potential \eqref{radeff} can be approximated as \eqref{effeik}. By considering the near-ring region and assuming that $\omega_R \gg \omega_I$, we can further specify \eqref{wkb} as
\begin{equation}\label{reso}
\frac{\alpha^2\omega^2-\frac{\lambda f(r_{0})}{r_{0}^{2}}}{f(r_{0})\sqrt{2V_{\lambda}^{\prime\prime}}}=-i\left(n+\frac{1}{2}\right).
\end{equation}

Taking the real part of the complex algebraic equation \eqref{reso}, we find
\begin{equation}
\alpha^2\omega_{R}^2-\frac{\lambda f(r_{0})}{r_{0}^{2}}=0,
\end{equation}
which gives us
\begin{equation}
\omega_R=\frac{\sqrt{\lambda f(r_{0})}}{r_{0}}.
\end{equation}
The imaginary part $\omega_I$ of the QNM frequency can be obtained by considering
\begin{equation}
2\alpha^2\omega _I \omega_R=\left(n+\frac{1}{2}\right) f(r_0)\sqrt{2V_{\lambda}^{\prime\prime}},
\end{equation}
and explicitly written as
\begin{equation}\label{omeii}
\omega_I=\frac{\sqrt{2}(2n+1)r_{0}\sqrt{f(r_{0})V_{\lambda}^{\prime\prime}}}{4 \alpha^2\sqrt{\lambda}}.
\end{equation}
As a result, we obtain the complex frequency of the QNMs in the eikonal limit as
\begin{equation}
\omega=\frac{\sqrt{\lambda f(r_{0})}}{r_{0}}-i \left(n+\frac{1}{2}\right)\frac{|\gamma_S|}{\alpha^2},
\end{equation}
where we have used the standard definition of the Lyapunov exponent from \cite{Cardoso:2008bp} as
\begin{equation}\label{lyawwsts}
\gamma_S=\frac{r_{0}\sqrt{f(r_{0})V_{\lambda}^{\prime\prime}}}{\sqrt{2\lambda}}.
\end{equation}
In the following, we will see that this Lyapunov exponent is related to the demagnification of the black hole images.

\subsubsection{\texorpdfstring{$\mathfrak{s l}(2, \mathbb{R})_{\mathrm{PR}}$}{lg3}  of null geodesics}

The radial momentum $P_r$ of the photon can be determined from the Lagrangian \eqref{lagc} as 
\begin{equation}
P_r =-\frac{\partial \mathcal{L}}{\partial \dot{r}}=\frac{1}{\Omega^2 f(r)}\cdot \frac{\mathrm{d}r}{\mathrm{d}\tau} \pm\frac{1}{f(r)}\sqrt{\alpha^2 E^2-\frac{K^2 S^2}{P(\theta_p) \sin ^2 \theta_p} \frac{f(r)}{r^2}} \equiv \pm \frac{\sqrt{\mathcal{V}(r)}}{f(r)}.
\end{equation}
We can introduce a coordinate transformation from $\{r, \phi, P_r, P_\phi\}$ to $\{T, \Phi, H, S\}$, where
\begin{align}
H&=P_t=\sqrt{-\frac{g^{11}}{g^{00}}P_r^2-\frac{g^{33}}{g^{00}}P_\phi^2}, \\
 \mathrm{d} T&=\frac{H}{f(r) \sqrt{\mathcal{V}(r)}} \mathrm{d} r, \\ 
  \mathrm{d} \Phi&=\mathrm{d} \phi-\frac{K^2 S}{r^2 P(\theta_p)\sin^2\theta_p \sqrt{\mathcal{V}(r)}} \mathrm{d} r.
 \end{align}
 It can be verified that the local canonical symplectic form
 \begin{equation}
 \bar{\Omega}=\mathrm{d} P_r \wedge \mathrm{d} r+\mathrm{d} P_\phi \wedge \mathrm{d} \phi=\mathrm{d} H \wedge \mathrm{d} T+\mathrm{d} S \wedge \mathrm{d} \Phi
 \end{equation}
 is preserved on the $\mathbb{R}^4$  phase space $\Gamma$.

The equations of motion for $H, S, T, \Phi$ are
\begin{equation}\label{hdotldot}
\dot{H}=\{H, H\}=0, \quad \dot{S}=\{S, H\}=0,
\end{equation}
\begin{equation}\label{tdoth}
\dot{T}=\{T, H\}=\frac{\partial T}{\partial r}\frac{\partial H}{\partial P_r}-\frac{\partial T}{\partial P_r}\frac{\partial H}{\partial r}=1,
\end{equation}
\begin{equation}\label{phidotph}
\dot{\Phi}=\{\Phi, H\}=\frac{\partial \Phi}{\partial \phi}\frac{\partial H}{\partial P_r}-\frac{\partial \Phi}{\partial P_r}\frac{\partial H}{\partial \Phi}=\frac{2H}{f(r)\sqrt{\mathcal{V}}}\left(1-\frac{K^2 S}{r^2 P\sin^2\theta_p \sqrt{\mathcal{V}(r)}} \frac{\mathrm{d}r}{\mathrm{d}\phi}\right) =0,
\end{equation}
where we denoted $\dot{\mathfrak{X}}={\mathrm{d}}\mathfrak{X}/{\mathrm{d}}T$. \eqref{hdotldot} confirms that $H$ and $S$ are conserved in the superselection sectors that $\Gamma$ is foliated into. \eqref{tdoth} and \eqref{phidotph} respectively yield
\begin{equation}
\Delta\phi=\int_{\phi_s}^{\phi_o} \mathrm{~d} \phi=\int_{r_s}^{r_o} \frac{K^2 S}{r^2 P(\theta_p)\sin^2\theta_p \sqrt{\mathcal{V}(r)}}\mathrm{d} r,
\end{equation}
\begin{equation}
T=\int_{r_s}^{r_o} \frac{H}{f(r) \sqrt{\mathcal{V}(r)}} \mathrm{d} r.
\end{equation}
This means that in an elapsed time $T$, a photon with fixed $H$ and $S$ follows a geodesic trajectory from $(r_s, \phi_s)$ to $(r_o, \phi_o)$.

We only consider light rays that begin and end at null infinity, with an impact parameter $\lambda$ greater than the critical value $\lambda_p$ defined in \eqref{lwwhwit}. For this purpose, we introduce the quantity
\begin{equation}\label{hat489r}
\hat{H} \equiv H-\frac{|S|}{\lambda_p}<0.
\end{equation}
The critical curve, which represents the boundary of the shadow or the circular photon orbit, corresponds to $\hat{H}=0$. Following the suggestion of \cite{Hadar:2022xag}, we define
\begin{equation}\label{hp0d}
H_{+}=\hat{H}, \quad H_0=-\hat{H} T, \quad H_{-}=\hat{H} T^2.
\end{equation}
It can be easily verified that the $\mathfrak{sl}(2, \mathbb{R})$ algebra
\begin{equation}\label{dk489}
\left[H_{\pm}, H_{\mp}\right]=\pm 2 H_0, \quad\left[H_0, H_{\pm}\right]= \mp H_{\pm}
\end{equation}
 is satisfied. We refer to this as the $\mathfrak{sl}(2, \mathbb{R})_{\mathrm{PR}}$ algebra. It should be noted that other constructions of the $\mathfrak{sl}(2, \mathbb{R})_{\mathrm{PR}}$ algebra are possible when considering the relation $\dot{\Phi}=\{\Phi, S\}=1$ \cite{Chen:2022fpl}.

The $\mathfrak{sl}(2, \mathbb{R})_{\mathrm{PR}}$ algebra, as discussed in references \cite{Raffaelli:2021gzh, Hadar:2022xag, Kapec:2022dvc, Chen:2022fpl}, is specifically applied to photons on the equatorial plane for black holes with $\mathbb{Z}_2$ symmetry. For the accelerating black hole, we can deduce from equation \eqref{macie38} that $0\leqslant mA<1/2$. Consequently, by referring to  \eqref{csothetap}, we can conclude that the $\mathfrak{sl}(2, \mathbb{R})_{\mathrm{PR}}$ algebra is valid for photons with $-1/3<\cos\theta_p\leqslant 0$. Additionally, choosing a reasonable range for the dimensionless parameter $mA$ results in distinct types of photon geodesics that adhere to the algebraic constraints.

In the case of a Schwarzschild black hole, the phase space is defined on a constant-$\theta$ hypersurface with $\theta=\pi/2$. However, for an accelerating black hole, the phase space is still on a constant-$\theta$ hypersurface, but with $\theta=\theta_p\neq\pi/2$. The generators of the $\mathfrak{sl}(2, \mathbb{R})_{\mathrm{PR}}$ algebra correspond to left-invariant points
\begin{equation}\label{lefinp}
\tilde{r}=r_p, \quad \tilde{\phi} \in[0,2 \pi), \quad \tilde{P}_r=0, \quad \tilde{P}_\phi= \pm \lambda_p H
\end{equation}
within the superselection sectors $\Gamma_S$ with fixed $S$, but variable photon energy $H$ and impact parameter $\lambda_p$. It should be noted that the homoclinic orbits cannot reach the pole points, so we do not have $\tilde{\phi} \in[0,2 \pi K)$. These points can be used to detect the finite dilation $\mathrm{e}^{-\alpha_0 H_0}$ generated by $H_0$ defined in \eqref{hp0d}. This gives
\begin{equation}
e^{-\alpha_0 H_0} \hat{H} e^{\alpha_0 H_0}=e^{-\alpha_0} \hat{H}.
\end{equation}
Here, $\alpha_0>0$ corresponds to a finite attracting dilation that causes the null geodesic to approach the left-invariant points \eqref{lefinp}. In the limit $\alpha_0\to \infty$, the geodesic approaches the bound orbit formed by the left-invariant points \eqref{lefinp}.

There is a discrete subgroup of $\mathrm{SL}(2, \mathbb{R})_{\mathrm{PR}}$ that acts successively on $\Gamma_S$, forming a discrete set of geodesics $\Gamma_o$. These geodesics can be labeled by the number of times $\omega=\Delta\phi/2\pi\gg 1$ they orbit the black hole, and the adjacent geodesics $\omega$ and $\omega\pm 1$ are related by the emergent discrete scaling symmetry $D_0=\mathrm{e}^{-2 \pi H_0}$. Geodesics in $\Gamma_o$ originate from a photon source outside the black hole and terminate at the observer's screen. On the observer's screen, the images of the photon source are demagnified and approach the critical curve by a factor $\mathrm{e}^{-\gamma_S \tau}=\mathrm{e}^{-\pi}$. Here, $\gamma_S$ is the Lyapunov exponent given by \eqref{lyawwsts}, and $\tau$ is the half-period of the photon ring at $r=r_p, \,\theta=\theta_p$. In other words, similar to the Schwarzschild case \cite{Hadar:2022xag}, the emergent conformal symmetry $\mathfrak{sl}(2, \mathbb{R})_{\mathrm{PR}}$ on the off-equatorial plane can also be reflected onto the image plane of the black hole.

%%%%%%%%%%%%%%%%%%%%%%%%%%%%%%%%%%%%%%%%%%%%%%%%%%%%
%%%%%%%%%%%%%%%%%%%%%%%%%%%%%%%%%%%%%%%%%%%%%%%%%%%%
%%%%%%%%%%%%%%%%%%%%%%%%%%%%%%%%%%%%%%%%%%%%%%%%%%%%
%%%%%%%%%%%%%%%%%%%%%%%%%%%%%%%%%%%%%%%%%%%%%%%%%%%%
%%%%%%%%%%%%%%%%%%%%%%%%%%%%%%%%%%%%%%%%%%%%%%%%%%%%

\subsection{Emergent  conformal symmetry for Taub-NUT black hole}\label{hctnbh}

In the previous section, we observed that the emergent conformal symmetry is preserved in the case of an accelerating spacetime, where cosmic strings connect the black hole to the conformal boundary. As a result, the black hole loses the $\mathbb{Z}_2$ symmetry that is characteristic of the Schwarzschild black hole. In this section, we will examine whether the hidden conformal symmetry still exists in the Taub-NUT black hole. In the Taub-NUT black hole, there are Misner strings that extend from the black hole to spatial infinity, causing the black hole to also lose its $\mathbb{Z}_2$ symmetry. Besides the gravitational mass, the black hole has the gravitomagnetic mass, or in other words, the Misner gravitational charges or the NUT charge. We will consider the Taub-NUT black hole in the AdS background, but the results hold true for the flat backgrounds as well. It is important to note that all symbols used in the following analysis are independent from those used in the previous section.

\subsubsection{Off-equatorial photon ring}

The metric of the Taub-NUT AdS black hole is given by
\begin{equation}\label{met2}
d s^2=-f(r)\left[\mathrm{d} t+2 n(\cos \theta+\mathbb{X}) \mathrm{d} \phi\right]^2+\frac{\mathrm{d} r^2}{f(r)}+(r^2+n^2) (\mathrm{d} \theta^2+\sin^2 \theta \mathrm{d} \phi^2),
\end{equation}
where $n$ is the gravitomagnetic mass parameter (NUT charge parameter, Misner charge parameter) and $|\mathbb{X}|\leqslant 1$ is the Manko-Ruiz parameter that adjusts the location of the string singularity. Specifically, when $\mathbb{X}=-1$, the Misner string is located only at the $\theta=0$ pole, and when $\mathbb{X}=1$, it is located at the $\theta=\pi$ pole. The blackening factor is given by
\begin{equation}
f(r)=\frac{1}{r^2+n^2}\left[r^2-n^2-2 M r+\frac{r^4+6 n^2 r^2-3 n^4}{l^2}\right],
\end{equation}
where $M$ and $l$ are the mass parameter and the AdS radius, respectively. The event horizon $r_+$ of the black hole is determined by the equation $f(r)=0$.

The Lagrangian for a null geodesic in the Taub-NUT AdS spacetime is given by
\begin{equation}
\mathcal{L}=\frac{1}{2} g_{\mu \nu} \dot{x}^{\mu} \dot{x}^{\nu}=0,
\end{equation}
where the dot represents the derivative with respect to the affine parameter $\tau_0$. From the Lagrangian, we can determine the conjugate momenta as
\begin{equation}
P_{\mu}=\frac{\partial \mathcal{L}}{\partial \dot{x}^{\mu}}=g_{\mu \nu} \dot{x}^{\nu}.
\end{equation}
As the spacetime described by \eqref{met2} possesses Killing vectors $\xi_t$ and $\xi_\phi$, we obtain the conserved orbital angular momentum and energy of the null geodesic as $J=P_\phi$ and $E=-P_t$, respectively. Furthermore, the Hamiltonian can be expressed as
\begin{equation}
H=P_{\mu} \dot{x}^{\mu}-\mathcal{L}=\frac{1}{2} g^{\mu \nu}P_\mu P_\nu.
\end{equation}

To derive the equation of motion for the null geodesic, we utilize the Hamilton-Jacobi equation
\begin{equation}\label{hj1}
\frac{\partial S}{\partial \lambda}=H=\frac{1}{2} g^{\mu \nu}P_\mu P_\nu,
\end{equation}
where the separated form of the Jacobian action $S$ is given by
\begin{equation}\label{hj2}
S=J \phi-E t+S_{r}(r)+S_{\theta}(\theta).
\end{equation}
By introducing a separation constant $K$, the expressions for $S_r$ and $S_\theta$ become
\begin{equation}
-f(r) \left(\frac{{\rm d}S_r (r)}{{\rm d}r}\right)^2=\frac{K}{n^2+r^2}-\frac{E^2}{f(r)},
\end{equation}
\begin{equation}
\left(\frac{{\rm d}S_\theta (\theta)}{{\rm d}\theta}\right)^2=K-\csc^2(\theta) \left(J+2 E n \cos (\theta) +2 E n \mathbb{X}\right)^2,
\end{equation}
for the functions $S_r$ and $S_\theta$. Consequently, the momenta of the null geodesic in the radial and latitudinal directions can be expressed as 
\begin{equation}\label{preqf}
P_r=\frac{\partial S}{\partial r}=f(r)^{-1}\sqrt{R(r)},
\end{equation}
\begin{equation}\label{prethf}
P_\theta=\frac{\partial S}{\partial \theta}=\sqrt{\Theta(\theta)},
\end{equation}
where the radial and latitudinal effective potentials are
\begin{equation}\label{ep1}
R(r)=E^2-\frac{K}{n^2+r^2}f(r),
\end{equation}
\begin{equation}\label{ep2}
\Theta(\theta)=K-\csc^2\theta  \left(J+2 E n \cos \theta +2 E n \mathbb{X}\right)^2.
\end{equation}

Therefore, utilizing \eqref{hj2}, \eqref{preqf}, and \eqref{prethf}, and following the approach in \cite{Carter:1968rr}, we can derive the geodesic equations as
\begin{align}
\int^{\theta} \frac{{\rm{d}} \theta}{\sqrt{\Theta}}&=\int^{r} \frac{{\rm{d}} r}{(n^2+r^2)\sqrt{R}},\\
\tau&=\int^{r} \frac{{\rm{d}} r}{\sqrt{R}},\\
t&=\int^{\theta} \frac{-2n\cot\theta\left(2nE\cot\theta+J\sin^{-1}\theta+2 E n \mathbb{X} \csc\theta\right) }{\sqrt{\Theta(\theta)}}\mathrm{d}\theta+\int^r \frac{E}{f(r)\sqrt{R(r)}}{\rm{d}}r,\\
\phi&=\int^\theta \frac{2nE\cot\theta+J\sin^{-1}\theta+2 E n \mathbb{X} \csc\theta}{\sqrt{\Theta(\theta)}\sin\theta}{\rm{d}}\theta.
\end{align}
Furthermore, after some calculations, we obtain the four-velocity of the null geodesics as \footnote{Different forms of the equations of motion can be found in \cite{Frost:2023enn}.}
\begin{align}
\frac{{\rm{d}}t}{{\rm{d}}\tau}&=\frac{-2n\cot\theta(2nE\cot\theta+J\sin^{-1}\theta+2 E n \mathbb{X} \csc\theta)}{n^2+r^2}+\frac{E}{f(r)},\label{ee318ac}\\
\frac{{\rm{d}}r}{{\rm{d}}\tau}&=\sqrt{R(r)},\\
\frac{{\rm{d}}\theta}{{\rm{d}}\tau}&=\frac{\sqrt{\Theta(\theta)}}{n^2+r^2},\\
\frac{{\rm{d}}\phi}{{\rm{d}}\tau}&=\frac{2nE\cot\theta+J\sin^{-1}\theta+2 E n \mathbb{X} \csc\theta}{(n^2+r^2)\sin\theta}.\label{ee318ac2}
\end{align}

Then, similar to the case of an accelerating black hole, we can determine the location of the photon ring around the Taub-NUT black hole by satisfying the conditions
\begin{equation}\label{poc3}
\Theta\left(\theta_p\right)=0, \quad \dot{\Theta}\left(\theta_p\right)=0, \quad \ddot{\Theta}\left(\theta_p\right)<0,
\end{equation}
\begin{equation}\label{poc4}
R\left(r_p\right)=0, \quad R^{\prime}\left(r_p\right)=0, \quad R^{\prime \prime}\left(r_p\right)>0.
\end{equation}
The separation constant $K_p$, angular momentum $J_p$, and angular position $\theta_p$ of the photon ring can be explicitly expressed as
\begin{equation}\label{kpval}
K_p=4 n^{2} \tan\theta_p^{2},
\end{equation}
\begin{equation}\label{jpval}
J_p=-2 n \sec\theta_p,
\end{equation}
\begin{equation}\label{btrotp}
\tan\theta_p=\frac{\sqrt{n^2+r_p^2}}{2 n \sqrt{f(r_p)}}.
\end{equation}
The impact parameter, defined as $\lambda_p=J_p/E_p$ with $E_p$ the energy of the photon on the photon ring, can be obtained from
\begin{equation}\label{nrps48}
(n^2+r_p^2)\sin^2\theta_p=f(r_p)\left[\lambda_p+2n\left(\cos\theta_p+\mathbb{X}\right)\right]^2.
\end{equation}
The radius of the photon ring is determined by the equation
\begin{equation}\label{tnadspr}
-f'(r_p) +\frac{2 r_p f(r_p)}{n^2+r_p^2}=0.
\end{equation}
For the vanishing cosmological constant case, the radial position of the photon ring is given by
\begin{equation}
r_p=\sqrt[3]{(M-i n) (M+i n)^2}+\frac{\left[(M-i n) (M+i n)^2\right]^{2/3}}{M+i n}+M,
\end{equation}
which reduces to $3M$ for the Schwarzschild black hole case.

%%%%%%%%%%%%%%%%%%%%%%%%%%%%%%%%%%%%%%%%%%%%%%%%%%%%
%%%%%%%%%%%%%%%%%%%%%%%%%%%%%%%%%%%%%%%%%%%%%%%%%%%%
%%%%%%%%%%%%%%%%%%%%%%%%%%%%%%%%%%%%%%%%%%%%%%%%%%%%
%%%%%%%%%%%%%%%%%%%%%%%%%%%%%%%%%%%%%%%%%%%%%%%%%%%%
%%%%%%%%%%%%%%%%%%%%%%%%%%%%%%%%%%%%%%%%%%%%%%%%%%%%

\subsubsection{\texorpdfstring{$\mathfrak{sl}(2, \mathbb{R})_{\mathrm{QNM}}$}{lg4} of massless scalar field}

Now we consider the propagation of the massless scalar field $\Phi (t,r, \theta, \phi)$ in the Taub-NUT AdS spacetime. The equation of motion for the massless scalar field is given by
\begin{equation}\label{kgeq}
\nabla^\alpha \nabla_\alpha \Phi=0.
\end{equation}
Due to the symmetry of the spacetime \eqref{met2}, we can decompose the scalar field as $\Phi(t, r, \theta, \phi)=e^{-i \omega t} R(r) Y(\theta, \phi)$, where $\omega=\omega_R-i\omega_I$ represents the QNM frequency. By doing so, we can obtain separate equations of motion for the radial and angular parts as
\begin{equation}\label{ratn2}
\begin{aligned} 
\left\{\partial_r\left[\left(r^2+n^2\right) f(r) \partial_r\right]+\left(r^2+n^2\right)\frac{\omega^2}{f(r)}+4 n^2 \omega^2-\chi\right\} R(r)=0,
\end{aligned}
\end{equation}
\begin{equation}\label{laefx57}
\left\{ -\frac{1}{\sin ^2 \theta}\left[\sin \theta \partial_\theta\left(\sin \theta \partial_\theta\right)+\left(\partial_\phi+i 2 n \omega(\cos \theta+\mathbb{X})\right)^2\right]  +4 n^2 \omega^2-\chi\right\} Y(\theta, \phi)=0,
\end{equation}
where $\chi$ is the separation constant.

After utilizing the substitution \cite{Kalamakis:2020aaj}
\begin{equation}
R(r)=\frac{1}{\sqrt{r^2+n^2}} Z(r),
\end{equation}
the radial equation \eqref{ratn2} transforms into
\begin{equation}\label{radi34}
f(r) Z^{\prime \prime}(r)+f^{\prime}(r) Z^{\prime}(r) +\left[\frac{\omega^2}{f(r)} h(r)^2-\frac{\chi}{r^2+n^2}-U_{\mathrm{TN}}(r)\right] Z(r)=0,
\end{equation}
where
\begin{align}
U_{\mathrm{TN}}(r)&=\frac{r f^{\prime}(r)}{r^2+n^2}+\frac{n^2 f(r)}{\left(r^2+n^2\right)^2},\\
h(r)^2&=1+\frac{4 n^2 f(r)}{r^2+n^2}.
\end{align}
Next, by introducing the tortoise coordinate $r_*$ defined as
\begin{equation}
\frac{\mathrm{d} r_*}{\mathrm{d} r}=\frac{h(r)}{f(r)},
\end{equation}
and further decomposing $Z(r)$ as $ \psi_{\chi \omega}=\sqrt{h(r)} Z(r)$, we finally obtain
\begin{equation}\label{effoutn}
\frac{\mathrm{d}^2}{\mathrm{d} r_*^2}  \psi_{\chi \omega}\left(r_*\right)+\left[\omega^2-\mathcal{U}_{\mathrm{TN}}\left(r_*\right)\right] \psi\left(r_*\right)=0,
\end{equation}
where the effective potential for the massless scalar field is given by
\begin{equation}
\mathcal{U}_{\mathrm{TN}}\left(r_*\right)=  \frac{f(r)}{h(r)^2}\left[ U_{\mathrm{TN}}(r)+\frac{\chi}{r^2+n^2}+\frac{1}{2} \frac{f^{\prime}(r) h^{\prime}(r)}{h(r)}+\frac{1}{2} \frac{f(r) h^{\prime \prime}(r)}{h(r)}-\frac{3}{4} \frac{f(r) h^{\prime}(r)^2}{h(r)^2}\right].
\end{equation}
This Schrödinger-like equation \eqref{effoutn} can be solved numerically with infalling boundary conditions at the horizon and Dirichlet boundary conditions at the spatial boundary. It should be noted that the angular equation should be solved first. Here, we do not need to solve the equation of motion.

After denoting 
\begin{equation}\label{eikeff}
\mathcal{Q}_{\chi\omega}\left(r_*(r)\right)=\omega ^2-\mathcal{U}_{\mathrm{TN}},
\end{equation}
\eqref{effoutn} 
can be rewritten as
\begin{equation}\label{chieffsch}
\frac{\mathrm{d}^2 \psi_{\chi \omega}}{\mathrm{d} r_*^2}+\mathcal{Q}_{\ell \omega}\left(r_*\right) \psi_{\chi \omega}=0.
\end{equation}
In the eikonal limit $\chi\gg 1$, the effective potential \eqref{eikeff} approximates to
\begin{equation}\label{utndkladjf34}
\mathcal{U}_{\mathrm{TN}}\to \mathcal{U}_{\chi} =  \frac{\chi f(r)}{h(r)^2 (r^2+n^2)}.
\end{equation}
The effective potential $\mathcal{U}_\chi$ in the eikonal limit vanishes at both the horizon and spatial infinity. Additionally, there exists an extremal point $r_0$ of $\mathcal{U}_\chi$ determined by $\mathrm{d}\mathcal{U}_\chi/\mathrm{d}r=0$, which can be equivalently expressed as 
\begin{equation}\label{fnfpdk48}
f(r_0)=\frac{(n^2+r^2_0)f^\prime (r_0)}{2 r_0}.
\end{equation}
Remarkably, this coincides with the location of the photon ring given by \eqref{tnadspr}. This implies that the effective potential $\mathcal{U}_\chi$ can be expanded near the photon ring as
\begin{equation}
\mathcal{Q}_{\chi \omega}\left(r_*\right) \approx  \mathcal{Q}_0(\chi, \omega, r_*(r_0))+\frac{1}{2} \mathcal{Q}_0^{(2)}(\chi, r_*(r_0))\left[r_*(r)-r_*(r_0)\right]^2,
\end{equation}
where
\begin{equation}
\mathcal{Q}_0^{(n)}\equiv\left(\frac{d^n \mathcal{Q}_{\chi \omega}\left(r_*\right)}{d r_*^n}\right)_{r_*(r_0)}.
\end{equation}
Similarly, following the same derivation from \eqref{wkb} to \eqref{omeii}, we obtain
\begin{equation}
\omega=\frac{\sqrt{\chi f(r_{0})}}{r_{0}}-i \left(n+\frac{1}{2}\right)|\gamma_J|,
\end{equation}
where we introduce the Lyapunov exponent as
\begin{equation}\label{gammastaub}
\gamma_J=\frac{r_{0}\sqrt{2f(r_{0})\mathcal{U}_{\chi}^{\prime\prime}}}{\sqrt{2\chi}}.
\end{equation}
By defining
\begin{align}
x & =\left[2 \mathcal{Q}_0^{(2)}(\chi)\right]^{1 / 4}\left[r_*-r_*(r_0)\right], \\
h(\chi, \omega) & =\frac{\mathcal{Q}_0(\chi, \omega)}{\sqrt{2 \mathcal{Q}_0^{(2)}(\chi)}},
\end{align}
the Schrödinger-like equation \eqref{chieffsch} takes on a new form 
\begin{equation}
\left(-\frac{\mathrm{d}^2}{\mathrm{d} x^2}-\frac{1}{4} x^2\right)  \psi_{\chi \omega}\left(r_*(x)\right)=h(\chi, \omega)  \psi_{\chi \omega}\left(r_*(x)\right).
\end{equation}
Finally, by defining operators $J_1,\,J_2,\,J_3, \, J_\pm$ as same as \eqref{jop1}-\eqref{jop4}, we can verify that the $\mathfrak{s l}(2, \mathbb{R})_{\mathrm{QNM}}$ algebra \eqref{sl2r} emerges again.

The $\mathfrak{s l}(2, \mathbb{R})_{\mathrm{QNM}}$ algebra that we discovered for the Taub-NUT black hole case possesses unique characteristics. First, the scalar field is minimally coupled with the black hole spacetime background, a deviation from the previous accelerating black hole case where the scalar field was conformally coupled with the background geometry. Second, unlike in the accelerating black hole case where we could construct a radial effective potential for the photon that closely resembles the radial effective potential of the scalar field, the radial effective potential \eqref{ep1} for the photon in this case appears quite distinct from the scalar field's \eqref{utndkladjf34}. However, fortunately, these two effective potentials reach their extrema at the same radial position, as determined by \eqref{fnfpdk48}. Furthermore, the latitudinal effective potential-which can be inferred from \eqref{laefx57}-is complex. To the best of our knowledge, the QNMs of the black hole have not been thoroughly studied yet, and we do not need to resolve the equation as this $\mathfrak{s l}(2, \mathbb{R})_{\mathrm{QNM}}$ is independent of the latitudinal behavior of the scalar field.

\subsubsection{\texorpdfstring{$\mathfrak{sl}(2, \mathbb{R})_{\mathrm{PR}}$}{lg1} of null geodesics}

The Killing-Yano tensor for the Taub-NUT AdS black hole \eqref{met2} is given by \cite{Kubiznak:2007kh,Rodriguez:2021hks}
\begin{equation}
\mathbf{k}= n \mathrm{d} r \wedge\left[\mathrm{d} t+2 n (\cos \theta+\mathbb{X})\mathrm{d}\phi\right]  +r (n^2+r^2) \sin \theta \mathrm{d}\theta \wedge d \phi.
\end{equation}
Using this, we can construct a Killing tensor $\bar{K}_{\alpha \beta}=k_{\alpha \kappa} k_\beta{ }^\kappa$ and the corresponding conserved quantity $K$ as a separation constant in \eqref{ep1} and \eqref{ep2}, given by $K=\bar{K}^{\alpha\beta}P_\alpha P_\beta$. Although the Taub-NUT AdS black hole does not have a rotation parameter $a$, the angular momentum of the black hole can be nonzero if there is an asymmetric distribution of the Misner strings, which corresponds to the case $\mathbb{X}\neq 0$ \cite{Hennigar:2019ive,Frodden:2021ces,Liu:2022wku,Yang:2023hll,Wu:2023fcw}. \footnote{It should be noted that there has been no investigation of the total angular momentum including the contribution of the asymmetric distribution of the Misner string for the AdS case, though \cite{Rodriguez:2021hks} discussed the AdS case with $\mathbb{X}=0$.} However, from \eqref{tnadspr} we can see that the non-zero angular momentum induced by the asymmetric distribution of the Misner strings does not lead to a separation of the retrograde and prograde photon ring orbits. They remain degenerate to the same orbit, as in the case of the Schwarzschild black hole. Nevertheless, due to the gravitomagnetic mass, the photon ring of the black hole deviates from the equatorial plane, i.e., $\theta_p\neq \pi/2$, as shown in \eqref{btrotp}.

Just as in the case of an accelerating black hole, we can choose the plane $\theta=\theta_p$ and perform a coordinate transformation from $\{r, \phi, P_r, P_\phi\}$ to $\{T, \Psi, H, J\}$. The Hamiltonian $H$ can be solved from the relation $g^{\mu\nu}P_\mu P_\nu=0$, which can be explicitly written as
\begin{equation}\label{fhlval}
g^{00} H^2+2 g^{03} P_\phi H+g^{11} P_r^2+g^{33} p_\phi^2=0.
\end{equation}
Here, we have used the condition $P_\theta=0$ from \eqref{prethf} and \eqref{poc3}. Furthermore, by defining
\begin{align}
{\rm{d}}\Phi &=\mathrm{d}\phi-\frac{{\rm{d}} r}{2(n^2+r^2)\sqrt{R}}, \label{dtdthetader}\\ 
 {\rm{d}} T &=\frac{E}{f(r)\sqrt{R(r)}}{\rm{d}}r,\label{dpsidthetadr}
 \end{align}
 we can verify that the coordinate transformation maintains the local canonical symplectic form
 \begin{equation}
 \bar{\Omega}=\mathrm{d} p_r \wedge \mathrm{d} r+\mathrm{d} p_\phi \wedge \phi=\mathrm{d} H \wedge \mathrm{d} T+\mathrm{d} J \wedge \mathrm{d} \Phi
 \end{equation}
 on the $\mathbb{R}^4$ phase space $\Gamma$. The equations of motion for $H, J, \Phi, T$ are given by
 \begin{equation}\label{hjkx}
 \dot{H}=\{H, H\}=0, \quad \dot{J}=\{J, H\}=0,
 \end{equation}
 \begin{equation}\label{psiphix}
 \dot{\Phi}=\{\Phi, H\}=0, \quad\dot{T}=\{T, H\}=1.
 \end{equation} 
 \eqref{hjkx} confirms that $H$ and $J$ are conserved quantities. \eqref{psiphix} corresponds to ${\rm{d}}T=0$ and ${\rm{d}}\Phi=0$ in \eqref{dtdthetader} and \eqref{dpsidthetadr}, respectively.

As shown above, the radial locus of the light ring in the Taub-NUT AdS black hole spacetime can be determined by \eqref{tnadspr}. The conserved quantities, $K$ and $J$, can be obtained via \eqref{kpval} and \eqref{jpval}, respectively. Then, using \eqref{fhlval}, we can find the zero-point energy $\tilde{H}$ for the circular bound orbit. By defining $\hat{H}=H-\tilde{H}$, we can describe the photon that is captured by the black hole with $\hat{H}>0$, while the unbound photon that begins and ends at spatial infinity has $\hat{H}<0$. Furthermore, we can define
\begin{equation}
H_{+}=\hat{H}, \quad H_0=-\hat{H} T, \quad H_{-}=\hat{H} T^2,
\end{equation}
and observe that $H_\pm$ and $H_0$ obey the $\mathfrak{s l}(2, \mathbb{R})_{\mathrm{PR}}$ algebra.

For the Taub-NUT black hole,  \eqref{btrotp} confirms that the $\mathfrak{sl}(2, \mathbb{R})_{\mathrm{PR}}$ algebra is valid for photons with $\theta_{\min}\leqslant\theta_p\leqslant \theta_{\max}$, where $\theta_{\min}\leqslant\pi/2$ and $\theta_{\max}\geqslant\pi/2$. The exact values cannot be explicitly given, but are determined concurrently by \eqref{btrotp} and \eqref{nrps48}. Another interesting feature of the $\mathfrak{sl}(2, \mathbb{R})_{\mathrm{PR}}$ algebra in this context is that, even with a non-zero angular momentum of the black hole in the case of $\mathbb{X}\neq 0$, the algebra can still be abstracted using the coordinate transformation from $\left\{r, \phi, P_r, P_\phi\right\}$ to $\{T, \Psi, H, J\}$. This is quite different from the Kerr black hole case in \cite{Hadar:2022xag}, where a more complex Hamiltonian flow is required.

The $\mathfrak{sl}(2,\mathbb{R})_{\mathrm{PR}}$ algebra acts transitively on the superselection sectors $\Gamma_J$ that foliate $\Gamma$, with fixed $J$ and $K$ but variable $H=\hat{H}+\tilde{H}$. Consequently, any two unbound geodesics in $\Gamma_J$ can be mapped to each other by the $\mathfrak{sl}(2,\mathbb{R})_{\mathrm{PR}}$ algebra. Similarly to the case of the accelerating black hole, on the off-equatorial plane $\theta=\theta_p$, there exist left invariant points $\tilde{r}=r_p$, $\tilde{\phi}\in[0,2\pi)$, $\tilde{P}_r=0$, and $\tilde{P}_\phi=\pm\lambda_p H$ within the superselection sectors $\Gamma_J$. In this case, the photon's angular momentum $J$ remains fixed, while the energy $H$ and impact parameter $\lambda_p$ vary, corresponding to the generators of the $\mathfrak{sl}(2,\mathbb{R})_{\mathrm{PR}}$ algebra. A discrete subgroup of $\mathrm{SL}(2,\mathbb{R})_{\mathrm{PR}}$ can act on two nearly bound null geodesics starting from and ending at spatial infinity in $\Gamma_J$ through a discrete scaling symmetry $D_0=e^{-2\pi H_0}$, transforming one geodesic with winding number $\omega\gg 1$ into another geodesic with winding number $\omega\pm 1$. Moreover, close to the critical curve on the observer's sky, the images of the Taub-NUT black hole are demagnified by $e^{-\gamma_J\tau}$, where $\gamma_J$ is given by  \eqref{gammastaub} and $\tau$ is the half-period of the photon on the photon ring, obtained using  \eqref{ee318ac}, \eqref{ee318ac}, \eqref{btrotp}, and \eqref{tnadspr}. Therefore, it is possible to observe the emergent conformal symmetry $\mathfrak{sl}(2,\mathbb{R})_{\mathrm{PR}}$ of the Taub-NUT black hole by studying its shadow.

\section{Notes on emergent conformal symmetry  near particle ring}\label{kjhj4857}

We will investigate the behavior of the nearly bound timelike geodesic and the emergent conformal symmetry of the massive scalar field around spherically symmetric black holes. Specifically, we will focus on studying the emergent conformal symmetry around the Schwarzschild black hole. We will denote the emergent conformal symmetry of the timelike geodesics and massive scalar field as $\mathfrak{s l}(2, \mathbb{R})_{\mathrm{MPR}}$ and $\mathfrak{s l}(2, \mathbb{R})_{\mathrm{MQNM}}$, respectively.

\subsection{$\mathfrak{s l}(2, \mathbb{R})_{\mathrm{MPR}}$ of timelike geodesics}

\begin{figure}[t!]
	\centering
	\includegraphics[width=5.5in]{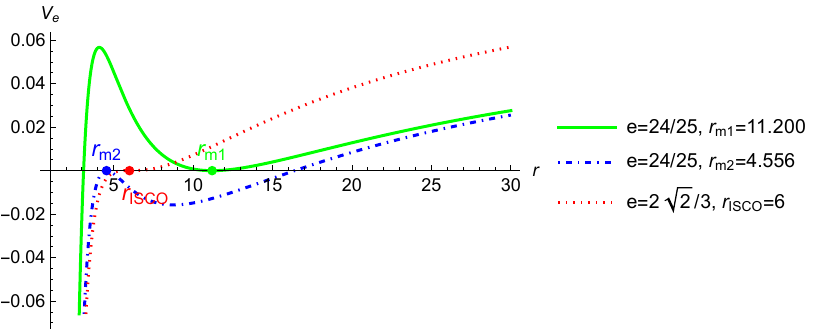}
	\caption{The effective potential of the massive particle is depicted in the figure. The green line represents the particle ring at $r_{m1}=11.200$, the dot-dashed blue line indicates the unstable particle ring at $r_{m2}=4.556$, and the dotted red line showcases the innermost stable circular orbit at $r_{\mathrm{ISCO}}=6$.}\label{effpar}
\end{figure}

The metric of the Schwarzschild black hole is given by
\begin{equation}
\mathrm{d} s^2=f(r) \mathrm{d} t^2-\frac{\mathrm{d} r^2}{f(r)}-r^2\left(\mathrm{d} \theta^2+\sin ^2 \theta \mathrm{d} \phi^2\right),
\end{equation}
where $f(r)=1-2M/r$. We will set the black hole mass $M=1$ in the following. The effective potential of the massive particle around the black hole, as derived in \cite{Chandrasekhar:1985kt}, is
\begin{equation}
V_e=f(r)\left(1+\frac{L^2}{r^2}-\frac{e^2}{f(r)}\right),
\end{equation}
where $e$ and $L$ represent the conserved energy and the angular momentum of the massive particle, respectively. See Fig. \ref{effpar} for specific examples of the effective potentials. The circular orbits of the massive particle (particle rings) are determined by the conditions
\begin{equation}
V_e\left(r_m\right)=V_e^{\prime}\left(r_m\right)=0,
\end{equation}
where ${}^\prime$ denotes the derivative with respect to the coordinate $r$, and $r_m$ denotes the radius of the circular orbit. These conditions yield the radius and the angular momentum for the stable and unstable circular orbits as
\begin{align}
r_{m1}&=\frac{-3 e^2+\sqrt{9 e^2-8} e+4}{2(1- e^2)},\,  L_1=\sqrt{\frac{27 e^4-36 e^2+8 \sqrt{9 e^2-8} e-9 \sqrt{9 e^2-8} e^3+8}{2 \left(e^2-1\right)}}\label{rmdkjr4},\\
r_{m2}&=\frac{3 e^2+\sqrt{9 e^2-8} e-4}{2 \left(e^2-1\right)},\,  L_2=\sqrt{\frac{27 e^4-36 e^2-8 \sqrt{9 e^2-8} e+9 \sqrt{9 e^2-8} e^3+8}{2 \left(e^2-1\right)}}\label{rmdkjr42}.
\end{align}
The range of the energy should be $2\sqrt{2}/3\leqslant e<1$. For $e=2\sqrt{2}/3$, the orbital radius and angular momentum for timelike geodesics (or the massive particle) on the innermost stable circular orbit are $r_{m1}=r_{m2}=6\equiv r_{\mathrm{ISCO}}, L_1=L_2=2\sqrt{3}\equiv L_{\mathrm{ISCO}}$. As $e\to 1$, we have $r_{m1}\to\infty, L_1\to \infty$, and $r_{m2}\to 4, L_2\to 4$. Generally, for $2\sqrt{2}/3< e<1$, we have $V_e^{\prime\prime}\left[r_{m1}\right]<0$, corresponding to the stable particle rings, and $V_e^{\prime\prime}\left[r_{m2}\right]<0$, corresponding to the unstable particle rings.

The Hamiltonian of the massive particle is given by \cite{Zhang:2019tzi}
\begin{equation}
2 \mathcal{H}=e \dot{t}-L \dot{\phi}-\frac{1}{f(r)} \dot{r}^2=1,
\end{equation}
where the dot denotes $\mathrm{d}/\mathrm{d}\tau$ with $\tau$ as the proper time. We can then introduce a coordinate transformation for the massive particle from $\left\{r, \phi, P_r, P_\phi\right\}$ to $\{T, \Phi, H, S\}$ as follows:
\begin{align}
H&=P_t=\sqrt{f(r)\left(1+\frac{P_\phi^2}{r^2}+f(r)P_r^2\right)}, \\
\mathrm{d} T&=\frac{H}{f(r) \sqrt{-V_e}} \mathrm{d} r, \\ 
\mathrm{d} \Phi&=\mathrm{d} \phi-\frac{L}{r^2\sqrt{-V_e}}\mathrm{d} r,
\end{align}
where 
\begin{equation}
P_r^2=-\frac{V_e}{f(r)^2}.
\end{equation}

For the massive particle, we can define an impact parameter $\lambda_m$ as
\begin{equation}
\lambda_m=\frac{L}{e}.
\end{equation}
This parameter, unlike the one for the photon, depends on the energy of the massive particle. Nevertheless, by following the same procedures from \eqref{hat489r} to \eqref{dk489}, we can show that an $\mathfrak{s l}(2, \mathbb{R})$ algebra still exists for the massive particle, whether on unstable or stable circular orbits. This $\mathfrak{s l}(2, \mathbb{R})_{\mathrm{MPR}}$ algebra for the massive particle may provide new insights into the shadow cast by massive particles \cite{Frost:2023enn,Kobialko:2023qzo}. That is, a discrete subgroup of $\mathrm{SL}(2, \mathbb{R})_{\mathrm{MPR}}$ can act on two nearly bound geodesics starting from and ending at spatial infinity through a discrete scaling symmetry $D_0=e^{-2 \pi (H-L/\lambda_m)T}$, transforming one geodesic into another adjacent one. Note that for the massive particle on the stable circular orbit $r_{m1}$, it has an imaginary Lyapunov exponent \cite{Cardoso:2008bp}
\begin{equation}
\gamma_P=\frac{i}{2}\sqrt{V_e^{\prime\prime}(r_{m1})\left(2-\frac{6}{r_{m1}}\right)}.
\end{equation}
This may result in long-lived QNMs around the stable circular orbit, similar to the case discussed in \cite{Guo:2021enm} for the photon and massless scalar field.

\subsection{$\mathfrak{s l}(2, \mathbb{R})_{\mathrm{MQNM}}$ of massive scalar field}

Now we turn our attention to the emergent conformal symmetry for the massive scalar field in the vicinity of a Schwarzschild black hole. We focus on the massive scalar field because the effective potential for the massless scalar field does not become extremal near the particle rings \eqref{rmdkjr4} and \eqref{rmdkjr42}. 

The Klein-Gordon wave equation for the massive scalar field with mass $\mu$ is given by
\begin{equation}
\nabla^\alpha \nabla_\alpha \Psi-\mu^2 \Psi=0,
\end{equation}
where the scalar field function can be decomposed as
\begin{equation}
\Psi=e^{-i \omega t} \frac{R(r, \omega)}{r} Y_{l m}(\theta).
\end{equation}
Here, $R$ and $Y$ are radial and angular functions respectively, describing the dynamics of the massive scalar field. With some calculations, we can derive a Schrödinger-like equation for the radial function
\begin{equation}
\frac{\mathrm{d}^2 R}{\mathrm{~d} y^2}+\left(\omega^2-V\right) R=0,
\end{equation}
where we have defined the tortoise coordinate by $dr/dy=f(r)$, and the explicit form of the effective potential $V$ is given by
\begin{equation}
V= f(r) \left(\mu ^2+\frac{\mathbb{K}}{r^2}+\frac{2 }{r^3}\right),
\end{equation}
where $\mathbb{K}=\ell (\ell+1)$ with $\ell\geqslant 0$ being the spherical harmonic index.

\begin{figure}[t!]
	\centering
	\includegraphics[width=5.5in]{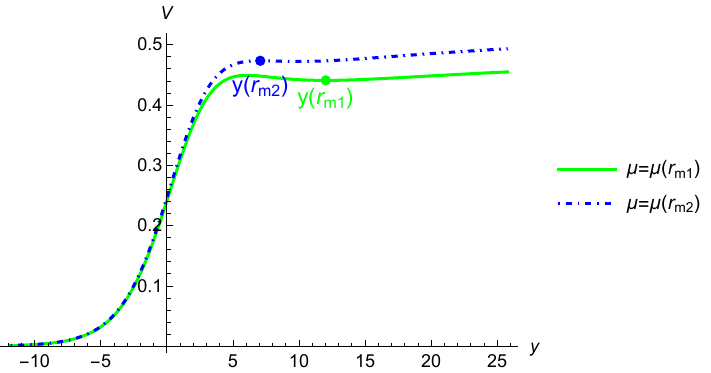}
	\caption{The effective potential of the massive scalar field for $\mathbb{K}=6$ exhibits interesting behavior. When the mass of the scalar field is related to the energy of the massive particle by the relation $\mu=\mu\left(r_{\mathrm{m} 1}\right)$, the effective potential has a local minimum at $y=y(r_{\mathrm{m1}})$. Conversely, when the mass of the scalar field is related to the energy of the massive particle by the relation $\mu=\mu\left(r_{\mathrm{m2} }\right)$, the effective potential has a local maximum at $y=y(r_{\mathrm{m2}})$.}\label{effpar84}
\end{figure}

We know that the effective potential of the massive scalar field can have extremal values at the locations $r_m=r_{m1}, r_{m2}$ of the circular orbits, provided that the mass of the scalar field fulfills the conditions
\begin{align}\label{jdk84dk}
V'(r_{m})=0, \quad\textrm{if}\quad \mu (r_m)=\frac{\mathbb{K} r_{m}^2-3 (\mathbb{K}-1) r_{m}-8}{r_{m}^3}.
\end{align}
With this in mind, let's examine the behavior of the massive scalar field near the stable/unstable massive circular orbit of the massive particle at $r_m=r_{m1}, r_{m2}$. After defining 
\begin{equation}
\mathcal{V}=\omega^2-V,
\end{equation}
we have 
\begin{equation}
\mathcal{V}\left(y\right) \approx \mathcal{V}\left(y\left(r_{m}\right)\right)+\frac{1}{2} \mathcal{V}_0^{(2)}\left(y\left(r_{m}\right)\right)\left[y(r)-y\left(r_{m}\right)\right]^2,
\end{equation}
where
\begin{equation}
\left.\mathcal{V}_0^{(2)} (y(r_{m}))\equiv \frac{\mathrm{d}^2 \mathcal{V}\left(y\right)}{\mathrm{d} y^2}\right|_{y(r)=y\left(r_{m}\right)}.
\end{equation}
From Fig. \ref{effpar84}, we can see that 
\begin{equation}
\mathcal{V}_0^{(2)} (y(r_{m1}))>0,\quad \mathcal{V}_0^{(2)}(y(r_{m2})) <0.
\end{equation}
Thus, we find that the effective potential at the unstable circular orbit $y(r_{m2})$ is an inverted harmonic oscillator potential, and the effective potential at the stable circular orbit  $y(r_{m1})$ is the harmonic oscillator potential. However, following the same procedures in \eqref{ghkd758}-\eqref{sl2r}, we can observe that the $\mathfrak{sl}(2, \mathbb{R})_{\mathrm{MQNM}}$ algebra emerges for the massive scalar field around both the stable/unstable circular orbits of the massive particle around the Schwarzschild black hole. Specifically, for the harmonic oscillator potential case, we have
\begin{equation}
\tilde{H}=-\frac{\mathrm{d}^2}{\mathrm{d} x^2}+\frac{1}{4} x^2,
\end{equation}
where
\begin{align}
x  =e^{i\frac{\pi}{4}}\left[-2 \mathcal{V}_0^{(2)}(y(r_{m2}))\right]^{1 / 4}\left[r_*-r_*(r_0)\right].
\end{align}
Then, by introducing the operators
\begin{align}
\tilde{J}_1 & =-\frac{i}{2}\left(x \frac{\mathrm{d}}{\mathrm{d} x}+\frac{1}{2}\right),\label{jop1} \\
\tilde{J}_2 & =\frac{i}{2}\left(\frac{\mathrm{d}^2}{\mathrm{d} x^2}+\frac{1}{4} x^2\right), \label{jop2}\\
\tilde{J}_3 & =\frac{i}{2}\left(\frac{\mathrm{d}^2}{\mathrm{d} x^2}-\frac{1}{4} x^2\right)=-\frac{i}{2} H, \label{jop3}\\
\tilde{J}_{ \pm}&= \pm i \tilde{J}_1-\tilde{J}_2,\label{jop4}
\end{align}
we can obtain the $\mathfrak{s l}(2, \mathbb{R})_{\mathrm{MQNM}}$ algebra
\begin{equation}\label{sl2r}
\left[\tilde{J}_{\pm}, \tilde{J}_{\mp}\right]=\mp 2 \tilde{J}_3, \quad\left[\tilde{J}_3, \tilde{J}_{ \pm}\right]= \mp \tilde{J}_{ \mp}.
\end{equation}

As shown in Fig. \ref{effpar84}, the effective potential of the massive scalar field, whose mass satisfies \eqref{jdk84dk}, vanishes at the horizon but diverges at the spatial infinity. This is different from the case discussed for the massless scalar field in \cite{Guo:2021enm}, though both encounter a local minimal extremal value. The WKB method cannot be used here to calculate the QNMs of the massive scalar field, which we will not investigate in detail in this paper. As the QNMs for the massive scalar field here are not solvable analytically, it seems challenging to construct a connection between the QNM frequencies for the massive scalar field and the representation of the $\mathrm{SL}(2, \mathbb{R})$ group explicitly.

\section{Closing remarks}\label{seccr}

It is not immediately clear whether the emergent near-ring conformal symmetry persists in black holes without $\mathbb{Z}_2$ symmetry. However, through investigating the motion of the massless scalar field and nearly bound null geodesics in black hole spacetimes with acceleration or gravitomagnetic mass, we have found that the emergent $\mathfrak{s l}(2, \mathbb{R})_{\mathrm{QNM}}$ and $\mathfrak{s l}(2, \mathbb{R})_{\mathrm{PR}}$ algebras still exist in the near-ring region in these cases. The absence of $\mathbb{Z}_2$ symmetry leads to photon rings that deviate from the equatorial planes of these black holes. This is significantly different from the previous investigation of Schwarzschild and Kerr black holes with north-south symmetry, as well as the lower-dimensional warped AdS$_3$ as the near-extremal Kerr geometry. The crucial observation in our study is that even outside the equatorial plane, the radial effective potential of the photon ring and the radial effective potential of the massless scalar field in the eikonal approximation almost coincide at the radial location where both potentials reach their extrema.

The discovery of the near-ring $\mathfrak{s l}(2, \mathbb{R})$ symmetry for the massless scalar field and nearly bound null geodesics in black holes provides a deeper understanding of the photon ring. This opens the possibility of defining a CFT on the photon ring, allowing for the application of classical physics of black holes in accordance with the holographic principle \cite{Raffaelli:2021gzh,Hadar:2022xag}. Our findings further support the universal existence of the near-ring $\mathfrak{s l}(2, \mathbb{R})$ algebra in the Plebanski-Demianski family of solutions in Einstein's gravitational theory.  \footnote{Certainly, the precondition is that there is a photon ring outside the black hole, see \cite{Yin:2023pao} (and references therein) for the photon ring topology of asymptotically flat, AdS, and dS spacetimes and see \cite{Junior:2021dyw} for the possibility of lack of a photon ring in the Melvin spacetime. It is worth noting that in the case of the three-dimensional Bañados-Teitelboim-Zanelli (BTZ) black hole, whether it is rotating or non-rotating \cite{Banados:1992wn}, charged or uncharged \cite{Martinez:1999qi}, or accelerating or non-accelerating \cite{Astorino:2011mw,Xu:2011vp,Arenas-Henriquez:2022www,Arenas-Henriquez:2023hur}, there is no presence of a photon ring.} Consequently, it demonstrates the feasibility of constructing a CFT on the photon ring. In terms of observations, the emergent $\mathfrak{s l}(2, \mathbb{R})_{\mathrm{PR}}$ symmetry provides constraints on the substructure around the photon ring. This symmetry can be utilized to measure the Lyapunov exponents, which are related to the $\mathfrak{s l}(2, \mathbb{R})_{\mathrm{QNM}}$ algebra governing the scalar perturbation of accelerating or NUTty black holes. By doing so, it becomes possible to constrain the acceleration \cite{Ashoorioon:2022zgu} or NUT charge acting as the gravitoelectric charge \cite{Mukherjee:2018dmm} of the black hole.

As an extension of the emergent conformal symmetry for the photon and the massless scalar field, we have also found that the $\mathfrak{s l}(2, \mathbb{R})_{\mathrm{MPR}}$ algebra and the $\mathfrak{s l}(2, \mathbb{R})_{\mathrm{MQNM}}$ algebra can be constructed for the massive particle and the massive scalar field around a Schwarzschild black hole. One unique characteristic of the particle ring is the existence of a stable circular orbit with an imaginary Lyapunov exponent. The $\mathfrak{s l}(2, \mathbb{R})_{\mathrm{MPR}}$ algebra for such a stable orbit should also be applicable to a stable photon ring around a black hole in the gravitational theory minimally coupled with a quasi-topological electromagnetic action \cite{Liu:2019rib,Wei:2023bgp}. Unlike the photon, the behavior of the massive particle is determined by both its angular momentum and energy. This results in the conformal algebra of the scalar field around the particle ring emerging only for a nonzero given mass, and this mass of the scalar field is related to the energy of the massive particle.

In the case of lower-dimensional warped three-dimensional AdS black holes, a spacetime isometry algebra $\mathfrak{sl}(2, \mathbb{R})_{\mathrm{ISO}}$ exists. This is equivalent to the $\mathfrak{s l}(2, \mathbb{R})_{\mathrm{QNM}}$ algebra in the near-ring region when acting on the eikonal massless scalar field. It would be interesting to explore whether a similar correspondence exists for the extremal near-horizon geometry of the accelerating black hole \cite{Astorino:2016xiy} and the Taub-NUT black hole \cite{Sadeghian:2018bli}. Furthermore, based on our investigations in this paper, we propose that the Love symmetry \cite{Charalambous:2021kcz,DeLuca:2021ite,Cvetic:2021vxa,BenAchour:2022uqo,Kehagias:2022ndy,Charalambous:2023jgq}, Ladder symmetry \cite{Hui:2021vcv,Berens:2022ebl}, and near-zone symmetry \cite{Hui:2022vbh} may be preserved by spacetimes without $\mathbb{Z}_2$ symmetry. However, these conjectures require specific verifications. Lastly, we suggest that studying the potential emergent $\mathfrak{s l}(2, \mathbb{R})$ algebra for the massive particle and the massive scalar field around the rotating black hole could be a future work. It would also be interesting to investigate the relation between the scalarization of the black hole and the emergent conformal symmetry of the massive scalar field around the black hole \cite{Simone:1991wn,Zhang:2022wzm}.

\acknowledgments

We would like to thank Hongbao Zhang for useful discussions. We are also deeply grateful to the editor and referee for their significant contributions.  M. Z. is supported by the National Natural Science Foundation of China (Grants No. 12365010, No. 12005080, and No. 12064018).  J. J.  is supported by the National Natural Science Foundation of China with Grant No. 12205014, the Guangdong Basic and Applied Research Foundation with Grant No. 2021A1515110913, and the Talents Introduction Foundation of Beijing Normal University with Grant No. 310432102. M. Z. is also supported by the Chinese Scholarship Council Scholarship (No. 202208360067).

\providecommand{\href}[2]{#2}\begingroup\raggedright\endgroup

\end{document}